\documentclass[conference]{IEEEtran}
\usepackage[multiple]{footmisc}
\usepackage{amsmath,amssymb,amsfonts,latexsym}
\usepackage{enumerate}
\usepackage[framemethod=tikz]{mdframed}
\usepackage{xspace}
\usepackage{epsf,picinpar}
\usepackage{varioref}
\usepackage{colortbl,multirow,hhline}
\usepackage{listings}
\usepackage{amssymb}
\usepackage{adjustbox}
\usepackage{colortbl,multirow,hhline}
\usepackage{algorithmic}
\usepackage{algorithm}
\usepackage{caption}
\usepackage[normalem]{ulem}
\usepackage{xcolor}
\usepackage{pifont}
\usepackage{xcolor,colortbl}
\usepackage[hyphens]{url}
\usepackage{balance}
\usepackage{graphicx, subfigure}
\usepackage{longtable}
\usepackage{lscape}
\usepackage{multirow}
\usepackage{listings}
\usepackage{framed}
\usepackage{morefloats}
\usepackage[T1]{fontenc}
\usepackage{array}
\usepackage{pdfpages}
\usepackage{fancybox}
\usepackage{amsmath}
\usepackage{flushend}
\usepackage{booktabs}
\usepackage{enumitem}
\usepackage{hyperref}
\usepackage{makecell}
\usepackage{graphics}

\definecolor{Grey}{gray}{0.9}

\usepackage[acronym]{glossaries}
\newacronym{llama}{LLaMA}{Large Language Model Meta AI}
\newacronym{cllama}{Code Llama}{Code Large Language Model Meta AI}
\newacronym{llm}{LLM}{large language model}
\newacronym{apr}{APR}{Automated Program Repair}
\newacronym{gqm}{GQM}{Goal Question Metric}
\newacronym{ai}{AI}{Artificial Intelligence}
\newacronym{nl2code}{NL2Code}{Natural-Language-to-Code}
\newacronym{clbg}{CLBG}{The Computer Language Benchmarks Game}
\newacronym{nby}{NBY}{n-body}
\newacronym{fkr}{FKR}{fannkuch-redux}
\newacronym{mlb}{MLB}{mandelbrot}
\newacronym{rpi}{Raspberry Pi}{Raspberry Pi 4 Model B}
\newacronym{cn}{CN}{Closest Numbers}
\newacronym{sr}{SR}{String Replacement}
\newacronym{ts}{TS}{Two Sum}

\newcommand{\CHANGE}[1]{\textcolor[HTML]{000000}{#1}}
\newcommand{\ie}{\emph{i.e.,}\xspace}
\newcommand{\eg}{\emph{e.g.,}\xspace}

\newcommand{\etal}{\emph{et~al.}\xspace} 

\makeatletter
\newcommand\notsotiny{\@setfontsize\notsotiny{6.4}{7}}
\makeatother

\begin{document}
\title{A Controlled Experiment on the Energy Efficiency \\of the Source Code Generated by Code Llama}

\author{\IEEEauthorblockN{Anonymous Authors}}

\author{
\IEEEauthorblockN{Vlad-Andrei Cursaru}
\IEEEauthorblockA{\textit{Vrije Universiteit Amsterdam}\\
The Netherlands\\
v.cursaru@vu.nl\\
}
\and
\IEEEauthorblockN{Laura Duits}
\IEEEauthorblockA{\textit{Vrije Universiteit Amsterdam}\\
The Netherlands\\
l.b.m.duits@student.vu.nl\\
}
\and
\IEEEauthorblockN{Joel Milligan}
\IEEEauthorblockA{\textit{Vrije Universiteit Amsterdam}\\
The Netherlands\\
j.a.milligan@student.vu.nl\\
}
\and
\IEEEauthorblockN{Damla Ural}
\IEEEauthorblockA{\textit{Vrije Universiteit Amsterdam}\\
The Netherlands\\
d.ural@vu.nl\\
}
\and
\IEEEauthorblockN{Berta Rodriguez Sanchez}
\IEEEauthorblockA{\textit{Vrije Universiteit Amsterdam}\\
The Netherlands\\
b.rodriguezsanchez@student.vu.nl\\
}
\and
\IEEEauthorblockN{Vincenzo Stoico}
\IEEEauthorblockA{\textit{Vrije Universiteit Amsterdam}\\
The Netherlands\\
v.stoico@vu.nl\\
}
\and
\IEEEauthorblockN{Ivano Malavolta}
\IEEEauthorblockA{\textit{Vrije Universiteit Amsterdam}\\
The Netherlands\\
i.malavolta@vu.nl\\
}}

\maketitle

\begin{abstract}
\noindent \textit{Context}. 
Nowadays, 83\% of software developers use Large Language Models (LLMs) to generate code. LLMs recently became essential to increase the productivity of software developers and decrease the time and cost of software development. Developers ranging from novices to experts use LLM tools not only to detect and patch bugs, but also to integrate generated code into their software. However, as of today there is no objective assessment of the energy efficiency of the source code generated by LLM tools. Released in August 2023, Code Llama is one of the most recent LLM tools.

\noindent \textit{Goal}. 
In this paper, we present an empirical study that assesses the energy efficiency of Code Llama with respect to human-written source code. 

\noindent \textit{Method}.
We design an experiment involving three human-written benchmarks implemented in C++, JavaScript, and Python. We ask Code Llama to generate the code of the benchmarks using different prompts and temperatures. Therefore, we execute both implementations and profile their energy efficiency.

\noindent \textit{Results}. 
Our study shows that the energy efficiency of code generated by Code Llama is heavily-dependent on the chosen programming language and the specific code problem at hand. Also, human implementations tend to be more energy efficient overall, with generated JavaScript code outperforming its human counterpart. Moreover, explicitly asking Code Llama to generate energy-efficient code results in an equal or worse energy efficiency, as well as using different temperatures seems not to affect the energy efficiency of generated code.

\noindent \textit{Conclusions}. 
According to our results, code generated using Code Llama does not guarantee energy efficiency, even when prompted to do so. Therefore, software developers should evaluate the energy efficiency of generated code before integrating it into the software system under development.
\end{abstract}

\section{Introduction}\label{c:introduction}
The usage of AI tools in the development process has increased rapidly over the last few years. The annual developer survey of 2023 conducted by Stack Overflow~\cite{Yepis_2023} found that, amongst 90,000 developers and technologists, 44\% are already using these AI tools. Moreover, the majority (77\%) is said to favour the usage of these tools in their workflow. 

There is a wide range of AI tools on the market today and the purpose of these tools varies from writing or debugging code to deploying and monitoring products. \acrfull{apr} techniques are, for instance, widely adopted to detect and provide patches for bugs~\cite{Huang_2023}. The majority of developers using AI tools (\ie 83\%) indicate to use these tools to generate code~\cite{Yepis_2023}. \textit{Large Language Models (LLMs)} are a specific category of generative AI tool that can assist developers write code. GitHub Copilot\footnote{\url{https://github.com/features/copilot/}} is an \acrshort{llm} model that generates code snippets based on the context provided by the user (\eg surrounding code and comments). Similarly, OpenAI’s 
ChatGPT\footnote{\url{https://chat.openai.com}} is able to translate natural language to code. On August 24, 2023, Meta AI released its own \acrshort{llm} model: Code Llama. Code Llama is a variant of Llama 2\footnote{\url{https://www.llama2.ai/}}, a general-purpose \acrshort{llm} model, obtained by training Llama 2 with code-specific datasets.

The training phase of \acrshort{llama} and Code \acrshort{llama} already produced about 1,015 tons of carbon emission (tCO2eq)~\cite{Touvron_2023} and 63.5 tCO2eq~\cite{Roziere_2023}, respectively. The developers of \acrshort{llama} hope that releasing the models will reduce greenhouse gas emissions in the future because, due to their small size, they can run on less powerful hardware (\eg a single GPU), and, at the same time, there is no need to retrain them ~\cite{Touvron_2023}. Nevertheless, the solutions generated by the \acrshort{llama} models may be energy-greedy and compromise the overall software quality, if integrated ~\cite{perry2022users}.
While research has been conducted to evaluate the security of source code generated by \acrshort{llm} models~\cite{Khoury_2023, Pearce_2021}, there is an existing research gap when it comes to the energy efficiency of generated code. 

Overall software energy consumption has significantly increased over the last decades~\cite{Wang_2022}.
In 2023, software is responsible for 4-5\% of the greenhouse gas emissions worldwide\footnote{\url{https://stateof.greensoftware.foundation/insights/software-emissions-are-equivalent-to-air-rail-shipping-combined/}}. Code design and writing influence software energy efficiency. Developers should, for instance, carefully consider which APIs to utilize in order to improve the energy efficiency of the whole hardware/software system.~\cite{Singh_2015}.

The \textbf{goal} of this paper is to assess the \textit{energy efficiency} of the source code generated using Code Llama. To this purpose, we design and execute an empirical study that compares the energy efficiency of generated code against human-written source code. We select three programming problems implemented in three programming languages: C++, JavaScript, and Python. Therefore, we ask Code Llama to generate a solution (\ie the source code) to the three problems in each language. 

Both the human-written and generated solutions are executed on an \acrshort{rpi}. Following the first comparison, we explicitly ask Code Llama to generate an energy-optimized version of the same problems and compare them to their corresponding human-written version.

To the best of our knowledge, this work represents \textit{the first empirical evaluation of the energy efficiency of code generated by Code Llama}. 
The \textbf{results} of our study show that the energy efficiency of code generated by Code Llama is heavily dependent on the chosen programming language and the specific code problem at hand. Also, human implementations tend to be more energy efficient overall, with generated JavaScript code outperforming its human counterpart. Moreover, explicitly asking Code Llama to generate energy-efficient code results in an equal or worse energy efficiency, as well as using different temperatures seems not to affect the energy efficiency of the generated code.

In summary, the main \textbf{contributions} of this paper are:
\begin{itemize}
    \item an empirical assessment of the energy efficiency of source code generated by Code Llama against human-written source code; 
    \item an in-depth discussion of the implications of obtained results;
    \item the replication package of the study including the prompts used to generate the code, the implementations of the programming problems, and the scripts for orchestrating the experiments, as well as raw data, and data analysis scripts~\cite{replication_package1}.
\end{itemize}

Our study offers valuable insights for developers looking to leverage the rapidly advancing generative \acrshort{llm}s in their software projects, by assessing whether Code \acrshort{llama} is able to produce energy-efficient solutions readily-available to developers.
\section{Experiment Definition}
\label{c:planning}
\label{c:rq}
The goal of this study is defined using the Goal-Question-Metric (GQM) framework \cite{Basili94}. Table 1 summarises the aim of this paper, which investigates the energy efficiency of programs generated by Meta’s Code Llama from the point of view of software developers in the context of Large Language Models. 

\begin{table}[h]
\centering
\footnotesize{
\centering
\caption{Goal description using the GQM framework}
\begin{tabular}{c|c}
\hline
\textbf{Analyze} &  Code generated by Code Llama \\
\hline
\textbf{for the purpose of}  & Evaluation \\
\hline
\textbf{with respect to their} & Energy Efficiency \\
\hline
\textbf{from the point of view} of  & Software Developers \\
\hline
\textbf{in the context of}  & Large Language Models \\
\hline
\end{tabular}
\label{goal}
}
\end{table}
\vspace{-2mm}

\noindent This goal is achieved by answering three research questions:

\noindent \textit{RQ1: How does the energy efficiency of the generated code compare against human implementations?}
Developers now have easy access to code-generating LLMs such as CoPilot, ChatGPT, and Code Llama, which might be used as a productivity supplement. However, there is \textit{no guarantee} that the code generated from these models is energy efficient. This question aims to learn the energy efficiency of Code Llama-generated code across several languages and algorithms \textit{without} asking the LLM to optimize for energy use. To develop a baseline to compare against, this paper utilises code challenges from popular websites for such problems: LeetCode\footnote{\url{https://leetcode.com/problemset/all/}} and HackerRank\footnote{\url{https://www.hackerrank.com/domains/algorithms}}. These algorithms have the benefit of being representative of simple issues that developers may try to solve with AI.

\noindent \textit{RQ2: Can Code Llama improve the energy efficiency of code with prompting?}
As shown in literature \cite{refining-chatgpt-generated-code, Khoury_2023}, other LLMs have proven able to resolve security vulnerabilities, repair bugs, and improve their code style \textit{if prompted to do so}. Similarly, this question investigates whether Code Llama can refactor programs to improve their energy efficiency when prompted. To this purpose, a comparison is performed between the energy efficiency of code that was generated with no specific focus on energy efficiency to that of programs created where an energy-efficient solution was specifically requested.

\noindent \textit{RQ3: How does energy efficiency differ between different solutions generated from the same prompt?}
In the paper \cite{Roziere_2023}, the \textit{temperature} parameter is included to introduce a level of randomness into each response generated by Code  Llama when presented with the same prompt. This parameter regulates the ability of the model to generate creative and novel outputs (\ie the randomness) \cite{Amazon_CodeLLama_2023}. Building upon this idea, this research question considers the difference in energy efficiency within code variations generated from the same initial prompt by Code Llama. By doing so, this study aims to investigate the extent to which inherent randomness of Code Llama influences the energy efficiency of the code it generates.

\section{Experiment Planning}%
\subsection{Subjects Selection}\label{c:subjects}
To evaluate the energy efficiency of code generated by Code Llama, we selected three programming problems and compared the solutions to the problems generated by Code Llama against the corresponding human implementation. The human-written implementations of the problems are chosen at random from those available on the Internet. The problems were based on three characteristics that can impact software energy usage: CPU-bound, memory-bound, and I/O-bound. 
Problems that are CPU-bound are heavily energy-consuming as they require the CPU to work harder over a longer period of time \cite{etinski2012understanding}.
One example of such a problem is \acrfull{cn}, which determines the pairs with the smallest absolute difference. The solution considered for \acrshort{cn} involves iterating over a list of numbers and, in every step, calculating the absolute difference of a different pair in the list. This task is time-consuming, especially if the initial list is unsorted. 
We picked one of the implementations of CN provided by HackerRank\footnote{https://www.hackerrank.com/challenges/closest-numbers/problem}.
The second problem considered is \acrfull{ts}, which requires the program to find the two elements in a given array of size $n$ that make the sum $(x + y) == k$ for a given number $k$. \acrshort{ts} is CPU-bound and it complexity depends on the implementation. The human implementation of TS is taken from LeetCode\footnote{\url{https://leetcode.com/problems/two-sum/solutions/4082514/98-21-hashmap-time-complexity-1-line-code/}}. 
As third problem, we selected a typical memory- and I/O-bound problem referred, in this paper, as \acrfull{sr}. SR executes three steps: (i) reading the content of a file, (ii) searching for a given string, and (iii) replacing it with a new one. The computational complexity of SR depends on the size of the input file. 
As for the other problems, we found the human implementation of SR, in Python and JavaScript, on gray literature\footnote{\url{https://www.geeksforgeeks.org/how-to-search-and-replace-text-in-a-file-in-python/}}\footnote{\url{https://attacomsian.com/blog/nodejs-replace-string-in-file}}. Whereas, the C++ version was implemented by master's degree students in computer science.

\begin{table}
\centering
\footnotesize{
    \captionsetup{justification=centering}
    \caption{Programming problems used as subjects in this study}
    \begin{tabular}{ p{0.04\linewidth} | p{0.20\linewidth} | p{0.33\linewidth} | p{0.20\linewidth} }
        \noalign{\hrule height 1pt}
        \textbf{ID} & \textbf{Name} & \textbf{Description} & \textbf{Characteristic} \\ \hline
        \acrshort{cn} & \acrlong{cn} & Determining the smallest absolute difference given a list of numbers. & CPU-bound \\
        \rowcolor{Grey} \acrshort{ts} & \acrlong{ts} & Given an input array and a given integer k, find two elements x and y that make $(x + y) == k$. & CPU-bound \\
       \acrshort{sr} & \acrlong{sr} & Replacing a single string within a specified file. & Memory \& I/O-bound \\ 
        \noalign{\hrule height 1pt}
    \end{tabular}
    \label{table:benchmarks}
   }
   \vspace{-4mm}
\end{table}



\subsection{Experimental Variables}
\label{c:vars_hyp}
Energy efficiency is the only dependent variable of our study and is necessary to answer all three research questions. Energy efficiency is measured as energy consumption in millijoules (mJ) per execution of a specific problem implementation. In this study, we vary the characteristics of each implementation of a problem, which represent our independent variables. The code of each programming problem can be generated by Code Llama or human-written. The \textit{source code} plays a significant role in determining scenarios where either Code Llama-generated or human-written code performs better in terms of energy efficiency. Further, we analyze the same problem written in three \textit{programming languages}: C++, JavaScript, and Python. These languages are chosen according to the results of Pereira et al.~\cite{Pereira2017} and exhibit different energy efficiency: high, medium, and low, respectively. Moreover, these languages are among the most popular ones\footnote{According to the \href{https://www.tiobe.com/tiobe-index/}{TIOBE index of October 2023}}. In this way, we can observe if the characteristics and the popularity of a programming language have an impact on the generated code, as well as if Code Llama can optimize energy efficiency better in specific programming languages. For RQ2 and RQ3, we analyze factors related to the code generation process. These factors include the \textit{prompt} provided to Code Llama and the \textit{temperature} used. As mentioned in Section \ref{c:rq}, the temperature parameter regulates the randomness of the generated solution. We evaluate three different temperature values: 0.6, 0.8, and 1.0, representing low to high randomness. Varying the temperature can reveal if the energy efficiency of the generated solution can be deterministic or accidental.


\subsection{Experimental Hypotheses}
\subsubsection{RQ1}
The code generated by Code Llama is compared to the equivalent human implementation based on the average energy consumption $\mu$. 
To represent the space of combinations considered, $opt$ ($\in P\ x\ B$) refers to the Cartesian product of the sets of programming languages ($P = \{ C++,\ JavaScript,\ Python \}$) and programming problems ($B = \{ \acrshort{cn},\ \acrshort{ts},\ \acrshort{sr} \}$), while $src$ represents the code source: Code Llama-generated or human-written. Hence, $H^{src,opt}$ establishes the effect of the source of the code on the energy efficiency of a specific programming problem implemented in a particular language. As a result, the null hypothesis tests whether the average energy consumption of generated code and human-written code across different characteristics of a problem implementation is equal:

\begin{equation}\label{h0:rq1}
    H_0^{src,opt} : \mu_{e_{llama}}^{opt} = \mu_{e_{human}}^{opt} 
\end{equation}


The alternative hypothesis analyses whether the mean energy consumption is higher for Code Llama-generated code than its human implementation.

\begin{equation}\label{ha:rq1}
    H_a^{src,opt} : \mu_{e_{llama}}^{opt} > \mu_{e_{human}}^{opt} 
\end{equation}

\subsubsection{RQ2}
As a specific prompt given to a \acrshort{llm} has proven to affect the generated code~\cite{refining-chatgpt-generated-code, Khoury_2023}, the experiment analyses whether requesting an energy efficient solution affects the solution provided by \acrshort{cllama}. In order to do so, the following null hypothesis is used:

\begin{equation}\label{h0:rq2}
    H_0^{prt,opt} : \mu_{e_{initial}}^{opt} = \mu_{e_{improved}}^{opt} 
\end{equation}

Where $prt$ represents the specific prompt utilised to generate the code from \acrshort{cllama}. Here, two options are considered: the $initial$ prompt versus the $improved$ version generated with a prompt that explicitly requests for an energy efficient implementation. To inspect whether the $improved$ version indeed improves the energy efficiency, the alternative hypothesis states that the mean measured energy consumption is lower for the code generated by the $improved$ prompt. 

\begin{equation}\label{ha:rq2}
    H_a^{prt,opt} : \mu_{e_{initial}}^{opt} > \mu_{e_{improved}}^{opt} 
\end{equation}

\subsubsection{RQ3} 
The temperature parameter may lead Code Llama in generating different solutions for a given problem fixing programming language and prompt. Therefore, we analyse whether this parameter can help optimizing the energy efficiency of generated code. The null hypothesis states that the temperature has no significant impact on energy efficiency. In particular, Equation \ref{h0:rq3} expresses that the mean measured energy consumption for the generated programs is equal for a combination of programming languages ($P$) and programming problems ($B$), regardless of the temperature ($tmp = \{ 0.6,\ 0.8,\ 1.0 \}$). 

\begin{equation}\label{h0:rq3}
    H_0^{tmp,opt} : \mu_{e_{tmp_1}}^{opt} = \mu_{e_{tmp_2}}^{opt} = ... = \mu_{e_{tmp_k}}^{opt} 
\end{equation}

Instead, the alternative hypothesis tests if there is at least one temperature vlaue that causes a significant fluctuation in the measured mean such that $\exists(i,j) \in \{0.6,\ 0.8,\ 1.0\} |\ \mu_{e_{tmp_i}}^{opt} \neq \mu_{e_{tmp_j}}^{opt}$. 

\begin{equation}\label{ha:rq3}
    H_a^{tmp,opt} : \mu_{e_{tmp_1}}^{opt} \neq \mu_{e_{tmp_2}}^{opt} \neq ... \neq \mu_{e_{tmp_k}}^{opt} 
\end{equation}

\subsection{Experiment Design}\label{c:design}
This study is designed as \textit{multiple factor, multiple treatment} (MF-MT) since it involves four different factors and multiple treatments for each factor. We analyze the energy efficiency of the generated code by Code Llama varying the characteristics of the programming problems, \ie problem type and programming language, as well as parameters related to the generation process, \ie temperature and prompt.
These aspects are the \textit{factors} of our experiment. As described in Section \ref{c:subjects}, we created a benchmark including the human implementation of three programming problems: \acrfull{cn}, \acrfull{ts}, and \acrfull{sr}, which are different. Indeed, CN and TS rely on the CPU for most of their computation, while SR on memory and disk. The chosen algorithm influences the resulting energy consumption, making it a factor in this study.
Moreover, as mentioned in Section \ref{c:vars_hyp}, we analyze the impact of the programming language on the energy efficiency of the code generated by Code Llama. This factor has three \textit{treatments}: C++, JavaScript, and Python.
The temperature parameter present in \acrshort{cllama} controls the level of randomness of the code that is generated. For this factor, the treatments are 0.6, 0.8, and 1.0 corresponding to low to high randomness.
Finally, we change the prompt supplied to \acrshort{cllama} to see whether the model is able to improve the energy efficiency of the resulting code. In particular, we use a prompt to generate the code for each problem (\ie TS, CN, and SR) and another one to get the energy-optimized version.
For example, assuming we want to generate a program that prints the string \texttt{"Hello World!"} in Python and its energy-optimized version. Therefore, we will supply Code Llama with the prompt: "Write a program that prints \textit{'Hello World!'} in Python" and its energy-optimized version: "Write an energy-efficient program that prints \textit{'Hello World!'} in Python".
The experiment follows a \textit{full-coverage factorial design} to ensure that we cover every combination of each factor. The total number of trials in our experiment is as follows:
\begin{align*}
& \textit{Number of trials = } \textit{3 programming languages } \times   \\
& \times \textit{3 temperatures} \times \textit{2 prompts} \times \textit{3 algorithms} = \textit{54}
\end{align*}
\section{Experiment Execution}
\label{c:execution}
\subsection{Experimental Setting}
\label{c:setting}
\begin{figure}
    \centering
    \includegraphics[width=0.8\linewidth]{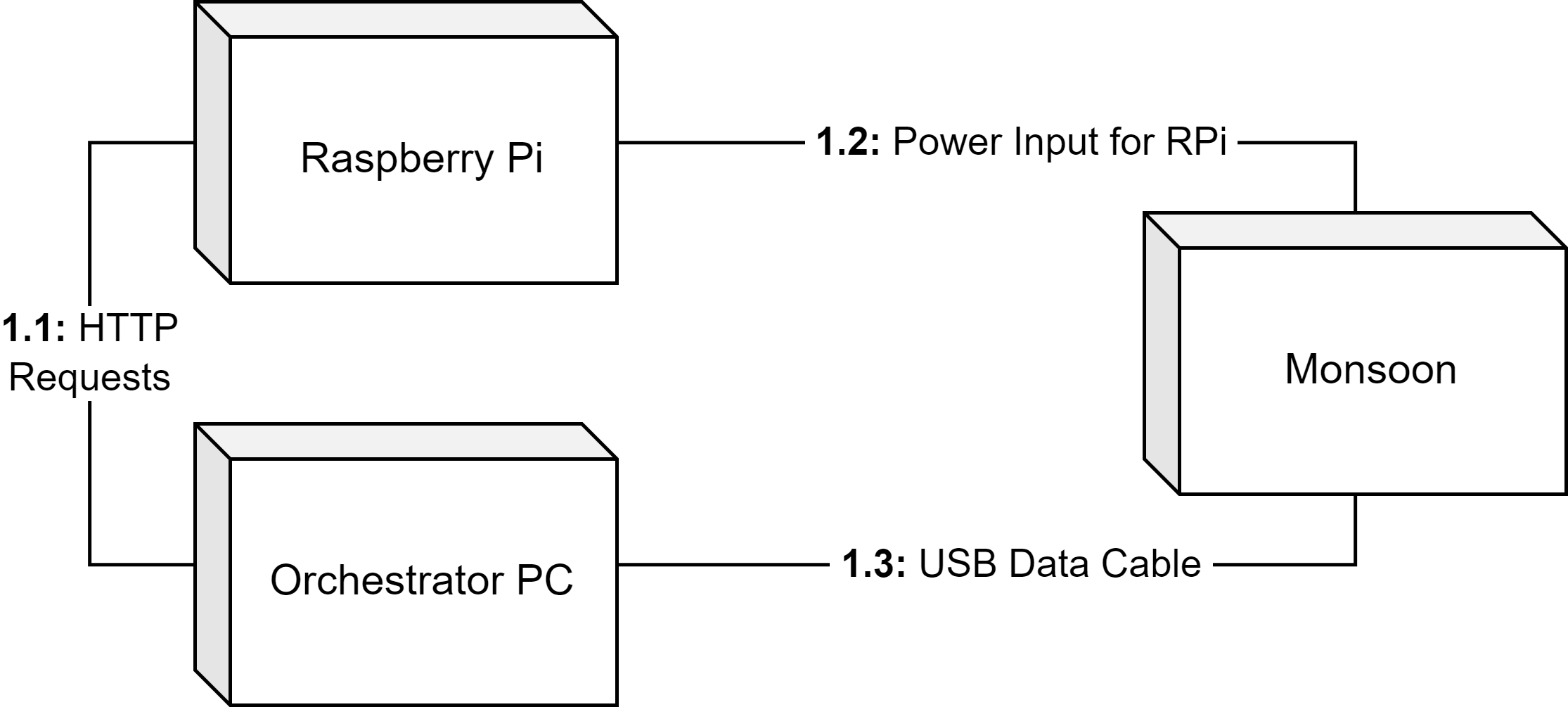}
    \captionsetup{justification=centering}
    \caption{Experimental Setting}
    \label{fig:monsoon-connection}
    \vspace{-4mm}
\end{figure}

We execute all the implementations of the programming problems on a \acrshort{rpi}, whereas we generate the code and, therefore, deploy Code Llama on a personal computer (PC) (\ie the orchestrator). The energy consumption of the \acrshort{rpi} is monitored using a Monsoon Power Monitor \cite{monsoon_2022}. The Monsoon Power Monitor is a physical power meter that measures the energy requested by the \acrshort{rpi} at a given frequency. The three devices are connected through USB 2.0 cables. Figure \ref{fig:monsoon-connection} shows the experimental setting used in this study. We decided to use a PC for code generation since it greatly outperforms the \acrshort{rpi}, speeding up the process. Instead, we use a \acrshort{rpi} for the energy measures due to its simpler architecture, which allows for a more controllable testbed. For example, due to the absence of a GPU on the \acrshort{rpi}, we did not consider any GPU-related optimizations for this experiment (this is left for future work).
The orchestrator includes a Ryzen 5 7600X processor, 32 GB of DDR5 RAM running at 5600 MHz, and an RTX 3060 12GB GPU. The code is generated using the \textit{CodeLlama-7b-instruct} model \cite{codeLLamaModel}, 
running on Ubuntu 22.04 WSL. We select the CodeLlama-7b-instruct model because of its compatibility with our hardware. Indeed, more powerful LLMs models require more computation resources, such as multiple GPUs \cite{zhang2022opt}. 
The orchestrator machine runs Experiment Runner \cite{experimentRunner}, which is a framework that facilitates setting up an experiment including pre-experiment actions, number of trials, and start/stop conditions for energy measures. The orchestrator issues commands to the \acrshort{rpi} via HTTP on a local network and gathers energy measures collected by the Monsoon Power Monitor via USB. 
The Monsoon Power Monitor connects to the \acrshort{rpi} via banana clips attached to ground and +5V pins of the \acrshort{rpi}. Before the execution of the experiment, we install on the \acrshort{rpi} all the tools and libraries required to run all the implementations of each programming problem. We use the following versions for each considered programming language:
\begin{itemize}
    \item \textbf{C++}: 20
    \item \textbf{Python}: 3.9.2
    \item \textbf{\CHANGE{JavaScript}}: 18.18.2 LTS
\end{itemize}
To avoid compatibility issues with the operating system and installed libraries on the \acrshort{rpi}, we compile the C++ code directly on the \acrshort{rpi} before execution.

As described in Section \ref{c:design}, we generate a total of 54 programs using the \texttt{src/instruction-runner.py} script available in the replication package \cite{replication_package1}.
The benchmark is generated using Code Llama in separate runs, which are based on the experimental factors. Consequently, the problem type, the programming language, the temperature value, and the prompt. 
The prompts used to generate the solutions for the programming problems are listed in \autoref{table:prompts}.

\begin{table}
\footnotesize{
    \captionsetup{justification=centering}
    \caption{Prompts supplied to Code Llama to generate the code for Closest Numbers (CN), Two Sum (TS), and String Replacement (SR).}
    \begin{tabular}{ p{0.1\linewidth} | p{0.75\linewidth} }
        \noalign{\hrule height 1pt}
        
        \textbf{ID} & \textbf{Prompt} \\ \hline
        
        \acrshort{cn} & Read a list of integers "arr" from standard input. Find the smallest absolute difference between any two elements of "arr". Return a list of tuples, only the pairs of elements of "arr" that have a difference equal to the smallest difference. \\
        
        \rowcolor{Grey} \acrshort{ts} & Take as arguments an array of integers arr and an integer k. Find two elements of arr, x and y, so that (x + y) == k. Print the indices of x and y in arr. \\
        
        \acrshort{sr} & Write a program that takes 3 arguments: a file, a first string and a second string. The program should replace all occurrences of the first string in the file with the second string. Consider that the files can be too large to fit in memory. Output the resulting text to a new file called out.txt. \\ 
        \noalign{\hrule height 1pt}
    \end{tabular}
    \label{table:prompts}
   }
   \vspace{-4mm}

\end{table}

\subsection{Data Collection}\label{c:baseline-data-collection}
As mentioned in Section \ref{c:setting}, we orchestrate the experiment using the Experiment Running framework, which configures the Monsoon Power Monitor, starts/stops recordings, executes the source code under analysis, and stores energy measures in a CSV file. 
These steps are repeated for all combinations of languages, algorithms, temperatures, and executions resulting in \( 3 \cdot 3 \cdot 3 \cdot 30 = 810 \) CSV files for Code Llama-generated code, and \( 3 \cdot 3 \cdot 30 = 270 \) CSVs for the human code. Therefore, these files are parsed and averaged to obtain the mean energy efficiency for the combinations of the aforementioned factors. The data collection process took a total of 36 hours. 
To answer RQ2, we investigate whether Code Llama can enhance the energy efficiency of its responses when specifically prompted to produce energy-efficient source code. Thus, we append to the prompts listed in Table \ref{table:prompts} the sentence: \textit{"Make the code as energy-efficient as possible"} to explicitly request Code Llama to generate an energy-efficient version of the source code.
After restructuring the prompts, we re-run the experiment. This resulted in 810  additional runs. After collecting the new energy measures, we statistically compare them with the results obtained from executing the code generated using the first version of the prompts.
\section{Results}\label{c:results}
This section presents the descriptive statistics and the results of the hypotheses testing (Section \ref{c:vars_hyp}) for each research question.
Table~\ref{tab:llama} provides an overview of the energy consumption measured for each problem (\ie \acrshort{cn}, \acrshort{ts}, and \acrshort{sr}) across prompts, programming languages, and temperatures. \autoref{tab:human-ds} includes the energy consumption measured for the respective human implementations. 
As each code is executed 30 times, a sample of 30 measurements is generated for each program. To test the normality of the data, we applied the Shapiro-Wilk normality test with a significance level of 0.05. The test results show that the data is non-normal. 


\begin{table*}
\centering
\notsotiny{
    \captionsetup{justification=centering}
    \caption{Descriptive statistics of the energy consumption (mJ) of Code \acrshort{llama}-generated code.}
    \begin{tabular}{|r|rrr|rrr|rrr||rrr|rrr|rrr|}
        \noalign{\hrule height 1pt}

        \multicolumn{19}{c}{\textbf{\acrlong{cn}}} \\ \noalign{\hrule height 1pt}

            \rowcolor{Grey}Prompt & \multicolumn{9}{c||}{Basic} & \multicolumn{9}{c|}{Efficient} \\ \noalign{\hrule height 1pt}
            
            \textbf{Language} & \multicolumn{3}{c|}{\textbf{C++}} & \multicolumn{3}{c|}{\textbf{JavaScript}} & \multicolumn{3}{c||}{\textbf{Python}} & \multicolumn{3}{c|}{\textbf{C++}} & \multicolumn{3}{c|}{\textbf{JavaScript}} & \multicolumn{3}{c|}{\textbf{Python}} \\ 
            
             Temperature 
                & 0.6 & 0.8 & 1.0 & 0.6 & 0.8 & 1.0 & 0.6 & 0.8 & 1.0 
                & 0.6 & 0.8 & 1.0 & 0.6 & 0.8 & 1.0 & 0.6 & 0.8 & 1.0 \\ \noalign{\hrule height 1pt}
    
            Mean 
                & 6.0 & 5.9 & 5.9 & 1.3 & 1.3 & 1.4 & 229.7 & 232.1 & 227.9
                & 4.0 & 6.3 & 6.3 & 3.4 & 3.4 & 3.4 & 237.3 & 232.6 & 230.0 \\
            Standard deviation 
                & 0.5 & 0.2 & 0.2 & 0.1 & 0.1 & 0.5 & 13.0 & 13.1 & 7.9
                & 0.8 & 0.1 & 0.1 & 0.1 & 0.3 & 0.2 & 43.9 & 17.0 & 10.9 \\ 
            Minimum 
                & 5.6 & 5.6 & 5.6 & 1.2 & 1.2 & 1.2 & 214.8 & 216.2 & 216.7
                & 3.6 & 6.1 & 6.1 & 3.2 & 3.1 & 3.2 & 215.0 & 211.8 & 214.8 \\
            Median 
                & 5.8 & 5.8 & 5.9 & 1.3 & 1.3 & 1.3 & 225.9 & 229.2 & 227. 2
                & 3.8 & 6.3 & 6.3 & 3.4 & 3.3 & 3.4 & 230.1 & 230.3 & 229.5 \\
            Maximum 
                & 7.7 & 6.2 & 6.3 & 1.8 & 1.6 & 4.1 & 270.2 & 287.7 & 248.3
                & 6.8 & 6.8 & 6.5 & 3.7 & 4.8 & 3.9 & 461.0 & 305.4 & 268.4 \\
            Coefficient of Variation 
                & 8.2 & 2.7 & 2.9 & 11.4 & 8.9 & 38.0 & 5.7 & 5.7 & 3.5 
                & 18.9 & 2.1 & 1.5 & 3.5 & 8.6 & 5.0 & 18.5 & 7.3 & 4.8 \\ \noalign{\hrule height 1pt}

        \multicolumn{19}{c}{\textbf{\acrlong{ts}}} \\ \noalign{\hrule height 1pt}

            \rowcolor{Grey}Prompt & \multicolumn{9}{c||}{Basic} & \multicolumn{9}{c|}{Efficient} \\ \noalign{\hrule height 1pt}
            
            \textbf{Language} & \multicolumn{3}{c|}{\textbf{C++}} & \multicolumn{3}{c|}{\textbf{JavaScript}} & \multicolumn{3}{c||}{\textbf{Python}} & \multicolumn{3}{c|}{\textbf{C++}} & \multicolumn{3}{c|}{\textbf{JavaScript}} & \multicolumn{3}{c|}{\textbf{Python}} \\ 
            
             Temperature & 0.6 & 0.8 & 1.0 & 0.6 & 0.8 & 1.0 & 0.6 & 0.8 & 1.0 & 0.6 & 0.8 & 1.0 & 0.6 & 0.8 & 1.0 & 0.6 & 0.8 & 1.0 \\ \noalign{\hrule height 1pt}
    
            Mean 
                & 41.4 & 35.6 & 35.7 & 1.2 & 1.2 & 1.1 & 367.5 & 401.6 & 397.8 
                & 0.1 & 0.2 & 0.2 & 1.3 & 1.3 & 1.3 & 432.3 & 434.2 & 435.4 \\
            Standard deviation 
                & 0.6 & 13.1 & 13.1 & 0.6 & 0.4 & 0.4 & 133.8 & 64.1 & 63.0
                & 0.0 & 0.1 & 0.1 & 0.1 & 0.1 & 0.2 & 9.3 & 10.1 & 34.6 \\ 
            Minimum 
                & 40.4 & 0.13 & 0.13 & 0.1 & 0.1 & 0.1 & 0.1 & 283.2 & 283.7
                & 0.1 & 0.1 & 0.1 & 1.1 & 1.1 & 1.1 & 423.3 & 424.3 & 423.6 \\
            Median
                & 41.4 & 41.5 & 41.7 & 1.2 & 1.2 & 1.2 & 426.5 & 427.8 & 427.0
                & 0.1 & 0.1 & 0.1 & 1.2 & 1.2 & 1.2 & 429.0 & 430.6 & 427.8 \\
            Maximum 
                & 43.2 & 43.6 & 43.5 & 3.5 & 1.5 & 1.6 & 451.1 & 476.7 & 472.6
                & 0.3 & 0.6 & 0.6 & 1.7 & 1.6 & 1.7 & 462.2 & 460.3 & 615.7 \\
            Coefficient of Variation 
                & 1.4 & 36.8 & 36.7 & 47.3 & 31.8 & 32.0 & 36.4 & 16.0 & 15.8 
                & 25.6 & 58.8 & 57.1 & 11.6 & 10.3 & 12.0 & 2.1 & 2.3 & 7.9 \\ \noalign{\hrule height 1pt}

        \multicolumn{19}{c}{\textbf{\acrlong{sr}}} \\ \noalign{\hrule height 1pt}

            \rowcolor{Grey}Prompt & \multicolumn{9}{c||}{Basic} & \multicolumn{9}{c|}{Efficient} \\ \noalign{\hrule height 1pt}
            
            \textbf{Language} & \multicolumn{3}{c|}{\textbf{C++}} & \multicolumn{3}{c|}{\textbf{JavaScript}} & \multicolumn{3}{c||}{\textbf{Python}} & \multicolumn{3}{c|}{\textbf{C++}} & \multicolumn{3}{c|}{\textbf{JavaScript}} & \multicolumn{3}{c|}{\textbf{Python}} \\ 
            
             Temperature & 0.6 & 0.8 & 1.0 & 0.6 & 0.8 & 1.0 & 0.6 & 0.8 & 1.0 & 0.6 & 0.8 & 1.0 & 0.6 & 0.8 & 1.0 & 0.6 & 0.8 & 1.0 \\ \noalign{\hrule height 1pt}
    
            Mean 
                & 8.3 & 7.7 & 7.7 & 12.6 & 12.7 & 12.6 & 14.3 & 12.0 & 11.9 
                & 7.7 & 7.7 & 7.7 & 12.6 & 12.6 & 12.6 & 20.2 & 20.1 & 34.3 \\
            Standard deviation 
                & 2.7 & 0.1 & 0.2 & 0.2 & 0.4 & 0.3 & 1.7 & 0.3 & 0.2  
                & 0.1 & 0.2 & 0.1 & 0.3 & 0.2 & 0.4 & 1.9 & 1.3 & 0.5 \\
            Minimum 
                & 7.5 & 7.5 & 7.5 & 12.2 & 12.0 & 12.0 & 13.5 & 11.6 & 11.6 
                & 7.5 & 7.5 & 7.5 & 11.9 & 12.0 & 12.0 & 19.6 & 19.7 & 33.8 \\
            Median
                & 7.7 & 7.7 & 7.7 & 12.6 & 12.6 & 12.6 & 14.0 & 11.9 & 11.9 
                & 7.7 & 7.7 & 7.6 & 12.6 & 12.6 & 12.6 & 19.9 & 19.9 & 34.2 \\
            Maximum 
                & 22.0 & 8.0 & 8.1 & 13.0 & 12.8 & 13.7 & 23.2 & 13.3 & 12.4 
                & 8.0 & 8.1 & 7.9 & 13.1 & 13.0 & 14.1 & 30.0 & 26.8 & 36.1 \\
            Coefficient of Variation 
                & 32.1 & 1.7 & 2.1 & 1.4 & 2.9 & 2.4 & 11.9 & 2.5 & 1.5 
                & 1.2 & 2.0 & 1.5 & 2.1 & 1.7 & 3.0 & 9.2 & 6.3 & 1.3 \\ \noalign{\hrule height 1pt}
        
    \end{tabular}
    \label{tab:llama}
}
\vspace{-2mm}
\end{table*}

\subsection{RQ1: How does the energy efficiency of the generated code compare
against human implementations?}

\begin{table*}
\centering
\footnotesize{
    \captionsetup{justification=centering}
    \caption{Descriptive statistics of the energy consumption (mJ) for the human implementations.}
    \begin{tabular}{|r||rrr||rrr||rrr|}
        \noalign{\hrule height 1pt}
        \rowcolor{Grey}Algorithm & \multicolumn{3}{c||}{\acrlong{cn}} & \multicolumn{3}{c||}{\acrlong{ts}} & \multicolumn{3}{c|}{\acrlong{sr}} \\ \noalign{\hrule height 1pt}
        
        \textbf{Language} & \textbf{C++} & \textbf{JavaScript} & \textbf{Python} & \textbf{C++} & \textbf{JavaScript} & \textbf{Python} & \textbf{C++} & \textbf{JavaScript} & \textbf{Python} \\ \noalign{\hrule height 1pt}

        Mean & 0.2 & 1.6 & 0.4 & 0.4 & 1.5 & 0.7 & 8.0 & 13.8 & 11.9 \\
        Standard deviation & 0.11 & 0.10 & 0.04 & 0.04 & 0.13 & 0.05 & 0.22 & 0.29 & 0.49 \\
        Min & 0.13 & 1.47 & 0.39 & 0.38 & 1.36 & 0.67 & 7.83 & 13.06 & 11.30 \\
        Median & 0.15 & 1.56 & 0.41 & 0.42 & 1.50 & 0.76 & 8.00 & 13.80 & 11.85 \\
        Max & 0.60 & 1.90 & 0.62 & 0.52 & 1.86 & 0.95 & 9.07 & 14.48 & 13.63 \\
        Coefficient of Variation & 60.8 & 6.2 & 10.4 & 9.1 & 8.6 & 7.3 & 2.7 & 2.1 & 4.1 \\ \noalign{\hrule height 1pt}
        
    \end{tabular}
    \label{tab:human-ds}
    }
\end{table*}
\begin{figure*}
    \centering
    \includegraphics[width=\linewidth]{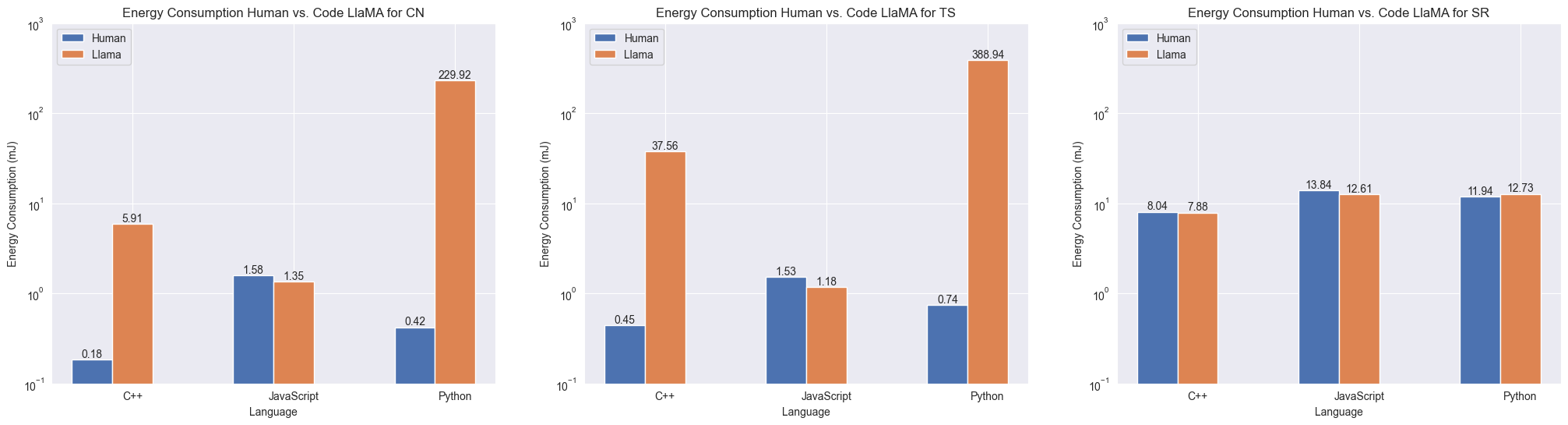}
    \captionsetup{justification=centering}
    \caption{Average energy consumption of human and generated code (in log scale).}
    \label{fig:rq1}
\end{figure*}

Figure~\ref{fig:rq1} compares energy consumption between generated and human code across different programming languages. As the energy consumption dataset presents a wide range of values, the y-axis in Figure~\ref{fig:rq1} is logarithmically scaled. To answer RQ1, we analyze the measurements obtained from executing the implementations generated using the prompts in Table \ref{table:prompts}, while varying temperatures. Implementations generated using energy-optimized prompts are excluded for this analysis.
In our experiment, Code Llama produced more energy efficient JavaScript code. Indeed, the generated JavaScript code consumed less energy than its human counterpart for CN, TS, and SR. The C++ and Python human implementations for CN and TS are more energy efficient than the generated ones.
There is no significant difference in energy consumption between different SR implementations. As opposed to CN and TS, the generated code outperformed the human code for C++ and JavaScript. On the contrary, the human implementation in Python proved to be more energy efficient.
Overall, the results of RQ1 indicate that the code generated by Code Llama tends to consume more energy with respect to the corresponding human implementation. However, the JavaScript implementations of the three problems generated by Code Llama turned out to be more energy-efficient. 

\subsubsection{Hypothesis Testing}
The Mann-Whitney non-parametric test is applied to compare two independent samples, which, in this case, are the energy consumption data from the human-written code and Code Llama-generated code. The p-values in \autoref{tab:pvalues-1} suggest that, for the problems and programming languages considered in this study, we can reject the null hypothesis. Thus, there is one group (\ie human-written or generated code) that has higher energy consumption values. 

\begin{table}[hbt!]
\centering
\footnotesize{
    \captionsetup{justification=centering}
    \caption{P-values for the Mann-Whitney test for RQ1.}
    \begin{tabular}{|l|ccc|}
        \noalign{\hrule height 1pt}
        \rowcolor{Grey} \textbf{Language}& \textbf{\acrshort{cn}} & \textbf{\acrshort{ts}} & \textbf{\acrshort{sr}} \\ \hline

        C++ & 1.69e-17 & 1.69e-17 & 2.74e-08\\
        JavaScript & 1.99e-10 & 7.50e-09 & 1.69e-17 \\
        Python & 1.69e-17 & 6.64e-09 & 7.61e-16\\ \noalign{\hrule height 1pt}
    \end{tabular}
    \label{tab:pvalues-1}
    }
\end{table}

Since we can reject the null hypothesis, we compute the effect size estimation with Cliff's delta \cite{macbeth2011cliff}. The Cliff's delta estimates are all close to 1 or -1. Therefore, there is a large effect size between generated code and human-written code. 

\subsection{RQ2: Can Code Llama improve the energy efficiency of code with prompting?}
\begin{figure*}
    \centering
    \includegraphics[width=1\linewidth]{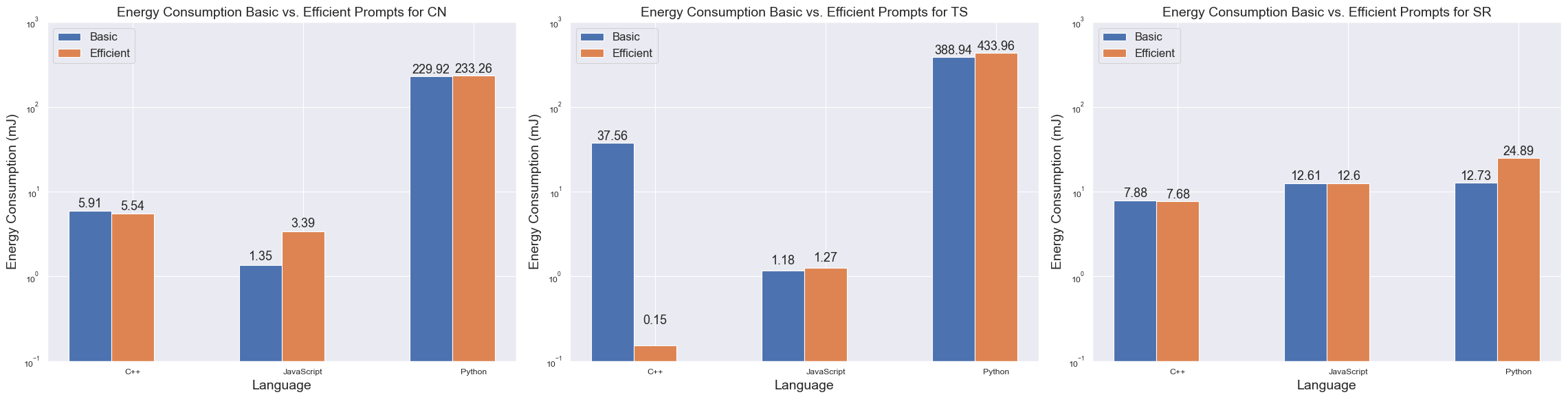}
    \captionsetup{justification=centering}
    \caption{Average energy consumption of code generated using basic and energy-efficient prompts (in log scale).}
    \label{fig:rq2}
\end{figure*}
To answer RQ2, we compared the energy consumption of the implementations generated with the initial prompt to those resulting from asking Code Llama an energy-efficient version of a problem implementation. The prompts are shown in Table \ref{table:prompts}. As described in Section \ref{c:baseline-data-collection}, we added the sentence: "Make the code as energy-efficient as possible" to ask Code Llama an energy-efficient version of the problems.
Table~\ref{tab:rq2} shows that the prompt has little impact on the implementation of CN. The maximum difference is obtained for Python (\ie 3.381 mJ). Figure~\ref{fig:rq2} shows more detailed results. Indeed, it shows that in \textit{most of the cases} the energy-optimized generated code \textit{consumes similar amount of energy as the original code}.
The results show that Code Llama is successful in generating energy-optimized code for the C++ implementations of CN and TS. In the latter, the difference is significant. However, since this is a single case this result could be accidental. In the other cases, the energy consumption variation is not significant. Overall the energy-efficient version of Python and JavaScript code results in a worse energy consumption.
\begin{table}[hbt!]
\centering
\footnotesize{
    \captionsetup{justification=centering}
    \caption{Difference in energy consumption between the initial optimized generated code.}
    \begin{tabular}{|l|ccc|}
        \noalign{\hrule height 1pt}
        \rowcolor{Grey}\textbf{Language} & \textbf{$\Delta$ CN} & \textbf{$\Delta$ TS} & \textbf{$\Delta$ SR} \\ \hline
        
        C++ & -0.374 & -37.408 & -0.2 \\ 
        JavaScript & 2.043 & 0.095 & -0.013 \\ 
        Python & 3.381 & 45.019 & 12.154 \\ \noalign{\hrule height 1pt}
    \end{tabular}
    \label{tab:rq2}
    }
\end{table}

\subsubsection{Hypothesis Testing}
The Mann-Whitney non-parametric test is applied to compare independent samples consisting of the energy consumption data for the original and the energy-optimized generated code. The p-values in \autoref{tab:pvalues-2} show that in some cases, such as for the TS prompt in JavaScript, there is not enough evidence to reject the null hypothesis. On the other hand, there are also cases, such as the CN prompt in JavaScript, where the null hypothesis can be rejected, which suggests that either the original or the improved version of the code had higher values for energy consumption. In this case, we do not compute the effect size estimation because there are instances where we could not reject the null hypothesis.

\begin{table}[hbt!]
\centering
\footnotesize{
    \captionsetup{justification=centering}
    \caption{P-values for the Mann-Whitney test for RQ2.}
    \begin{tabular}{|l|ccc|}
        \noalign{\hrule height 1pt}
        \rowcolor{Grey} \textbf{Language} & \textbf{\acrshort{cn}} & \textbf{\acrshort{ts}} & \textbf{\acrshort{sr}} \\ \hline
        
        C++ & 5.80e-11 & 1.69e-17 & 0.46 \\
        JavaScript & 1.69e-17 & 0.84 & 0.68 \\
        Python & 0.38 & 3.91e-03 & 3.89e-13\\ \noalign{\hrule height 1pt}
    \end{tabular}
    \label{tab:pvalues-2}
    }
\end{table}
\subsection{RQ3: How does energy efficiency differ between different solutions generated from the same prompt?}
\begin{figure*}
    \centering
    \includegraphics[width=1\linewidth]{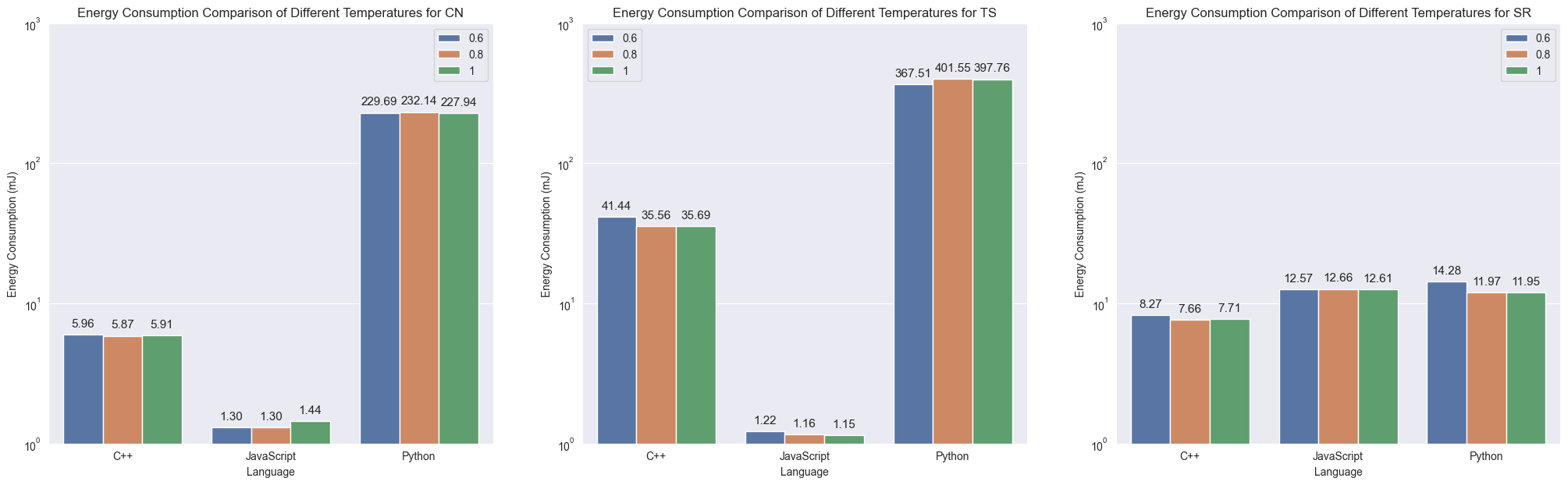}
    \captionsetup{justification=centering}
    \caption{Average energy consumption of code generated using different temperature values (in log scale).}
    \label{fig:rq3}
\end{figure*}

Each plot in Figure~\ref{fig:rq3} represents the energy consumption of each programming problem varying temperatures and programming languages. \textit{We observe that the temperature does not have a significant effect on the energy consumption}. Specifically, for CN, we observe that having different temperatures does not change the energy consumption significantly. The results for TS show that there is not a significant difference for JavaScript and Python. However, we observe that the energy consumption at 0.6 temperature for C++ is slightly higher compared to the other two temperatures. For SR, Python performed slightly worse at 0.6 than the other temperatures, while JavaScript and C++ exhibit similar values.

\subsubsection{Hypothesis Testing}
We applied the Kruskal-Wallis non-parametric test. It determines whether there are statistically significant differences between the energy consumption at different temperature values: 0.6, 0.8, and 1.0. The p-values in \autoref{tab:pvalues-3} reveal that only the SR prompt in Python had enough evidence to reject the null hypothesis, clearly seen on \autoref{fig:rq3}. For all other cases, the p-values prove that there is not enough evidence to reject the null hypothesis. We do not compute the effect size estimation because, in most cases, we could not reject the null hypothesis.
\begin{table}[hbt!]
\centering
\footnotesize{
    \captionsetup{justification=centering}
    \caption{P-values for the Kruskall-Wallis test for RQ3.}
    \begin{tabular}{|l|ccc|}
        \noalign{\hrule height 1pt}

        \rowcolor{Grey} \textbf{Language} & \textbf{\acrshort{cn}} & \textbf{\acrshort{ts}} & \textbf{\acrshort{sr}} \\ \hline
        C++ & 0.34 & 0.62 & 0.34 \\
        JavaScript & 0.51 & 0.80 & 0.78 \\
        Python & 0.31 & 0.17 & 1.27e-13 \\
        \noalign{\hrule height 1pt}
    \end{tabular}
    \label{tab:pvalues-3}
    }
    \vspace{-0.5mm}
\end{table}
\section{Discussion}\label{c:discussion}
\subsection{Reflections on the obtained results}
To provide for a meaningful interpretation of the data, we analyse the results presented in Section \ref{c:results} according to our three pairs of hypotheses. The first research question (\textit{RQ1}) focuses on the comparison between code generated by Code Llama, and human implementations. While there appears to be a significant difference between the energy efficiency of code generated by Code Llama and the code implementations provided by humans, neither consistently outperforms the other. Moreover, we reported similar findings for \textit{RQ2}: when comparing the energy consumption difference between the initial Code Llama implementation and the explicitly requested energy-efficient solution, Code Llama's ability to effectively generate efficient code varies on a case-by-case basis, depending on the language and algorithm, as well as the used prompt. We speculate that this could be attributed to the quality and quantity of code examples for each language present in the model training data. The model is trained ``predominantly on a near-deduplicated dataset of publicly available code''~\cite{Roziere_2023}, including questions and answers from forums. Such public code repositories are unlikely to enforce strict quality requirements on the code being shared, which we believe is likely to affect the outputs of the model as well.

Code Llama provides the possibility of adjusting the temperature of the models. However, from the results of \textit{RQ3}, it appears that this only affects the solutions generated for the \acrshort{sr} problem in Python. More specifically, the solution generated by a temperature of 0.6 deviates from the other two. The lower temperature means that the model stays closer to the prompt when generating a solution. Therefore, this result indicates that the \acrshort{sr} prompt may be misleading the model. The prompt explicitly states ``consider that the files can be too large to fit in memory'', which might be too specific, thus resulting in an inefficient implementation of the problem. 

In general, the Python solutions resulted in higher energy consumption. This is in line with the results from Pereira \etal \cite{Pereira2017}, who ranked Python as one of the most energy-consuming languages. Nevertheless, there is a significant difference between the energy efficiency of solutions generated by Code Llama versus the solutions provided by humans. Therefore, the energy consumption cannot be fully attributed to the programming language, but also partially to the model. Meta also provides a version of Code Llama specifically for Python, which could potentially result in more energy-efficient solutions. This is an avenue for further research. 

Figure \ref{fig:rq1} shows that the JavaScript code generated by Code Llama is more efficient for all the problems considered. This may depend on how Code Llama was trained, which may be more informed about specific types of languages and coding style. Indeed, the problems utilised in this study are either CPU-bound, I/O-bound and memory-bound. These characteristics can influence the results as the energy efficiency depends on the coding style, and therefore how the hardware is used by the software. For example, devices with less efficient CPUs will be affected more strongly by an inefficient CPU-bound workload. Similarly, slower memory may increase the runtime of the program in memory-intensive benchmarks and, as such, the overall power consumption. Furthermore, the results in Table \ref{tab:rq2} show that explicitly asking Code Llama to generate energy-efficient code does not bring any optimization in the generated code. \textit{This observation may suggest that Code Llama is unaware of tactics to improve the energy consumption of the code it generates}. For example, Code Llama could suggest ways to minimize unnecessary operations, introduce threads to parallelize independent tasks, and reduce I/O operations. Code Llama generates code that implements the desired functionality, but, considering our results, it does not appear to be able to generate energy-efficient code or optimize the energy consumption of the code it generates. In some cases, generated code is even worse than human-written code. Further experiments are needed to clarify this aspect.

\subsection{Implications for developers and organizations}
According to the results of our experiment, \textbf{it is important that developers are aware that the source code generated by Code \acrshort{llama} might not be as energy-efficient as human-generated source code}. While the models can generate correct code, this research shows that the energy efficiency of the resulting programs is inconsistent across different languages and tasks. As such, taking the generated code at face value may provide sub-optimal results. We would like to stress that we are not suggesting avoiding using LLMs altogether for development purposes. Nevertheless, \textbf{we urge developers to adopt a critical mindset regarding the efficiency of LLM-generated source code and its potential runtime impact}, encompassing not only energy efficiency but also other quality attributes such as performance, security, and reliability. 
In software projects where energy efficiency is of prime importance, LLM-generated source code can be used as a starting point for carrying out a programming task in an initial prototypical phase of the software product being realized; such source code can then be optimized via further iterations, at best after an empirical assessment of its energy efficiency.     

Given the widespread traction that LLMs are gaining as tools for supporting development activities~\cite{Yepis_2023}, \textbf{we invite Meta and the organizations maintaining other LLMs (\eg OpenAI and Google) to promote energy efficiency as a first-class feature to be considered during the training phases of their LLMs}. Producing energy-efficient code is not longer a desiderata, but it is rather a central quality aspect for the sustainability (and survival) of current and future software projects~\cite{verdecchia2021green}.
As an example, the European Union is actively working towards promoting and adopting measures towards achieving climate-neutral datacenters and communication networks and services by 2030, with an explicit mention of including ``software-related efforts such as intelligent energy-saving functions'', in addition to the various improvements at the hardware and infrastructural levels~\cite{bilsen2022study}. By prioritizing energy efficiency as a fundamental aspect in their reasoning processes when recommending source code to developers, LLMs can significantly contribute to turning this strategic vision into reality.
\section{Threats To Validity}\label{c:threats}
The aim of our study was to investigate the energy efficiency of Code Llama-generated code. In this section, we present the threats to validity that our study poses, divided into the four categories as defined by Cook and Campbell~\cite{Cook:1979}: \nameref{threat:internal}, \nameref{threat:external}, \nameref{threat:construct}, and \nameref{threat:conclusion}.

\subsection{Internal Validity}\label{threat:internal}
When running code, the amount of power used can differ significantly between different time stamps. That is why the choice was made to retrieve 5,000 measurements per second. This way, it would be possible to establish more accurate results for the energy consumed per run of a given solution. Nevertheless, these results may still fluctuate per run. Therefore, this threat to interval validity was mitigated by utilising the mean measured energy consumption of 30 runs. Additionally, in order to analyse the energy efficiency of the solutions for the given benchmarks, a Monsoon Power Monitor was used to measure the power usage of running the particular code. However, since the notebook might run additional processes in the background, this could affect the measurements. Therefore, we chose to run the code on a \acrshort{rpi} to mitigate this threat. Nonetheless, when a lot of code is run consecutively, this also affects the measured power usage. More specifically, the temperature accumulated between the various runs can affect the results. Since a cool-down period of 1 second between executions was used, this poses a threat to internal validity.  
Moreover, the Monsoon device experiences a limitation with respect to the conversion speed of the data frames to a CSV file. This results in dropped frames; \ie frames are thrown away if the Monsoon is not able to handle them fast enough. For some of the measurements, the percentage of dropped frames reached 1\% while for others it was in the range of 0\%. However, since the percentages are so small, we consider them negligible. Finally, picking up human-written code randomly from the internet can result in optimized code that is more energy efficient and, therefore, better results for the human-written code group. Despite this threat, Code Llama was able to generate more efficient code in some cases.

\subsection{External Validity}\label{threat:external}
The external validity of the experiment, as proposed and conducted in this paper, is threatened by various factors. Firstly, the choice for the three problems used as benchmarks in this study. While these problems were selected based on specific characteristics, two of the problems (\ie \acrshort{cn} and \acrshort{ts}) are CPU-bound, leaving \acrshort{sr} to represent both memory- and I/O-bound problems. However, due to the simplicity of the nature of the problems, the benchmarks are considered representative of the types of requests made by (novice) software developers to Code Llama. Moreover, the choice was made to employ three distinct programming languages in analysing the energy efficiency of Code Llama-generated code. The specific languages (\ie C++, JavaScript, and Python) were selected based on their energy-efficiency level and popularity. While the respective languages represent the high, middle, and low energy-efficient languages as found by~\cite{Pereira2017}, the popularity factor limits the generalisability of the findings. There is more data available on these popular languages in comparison to lesser-known languages such as TypeScript\footnote{According to the Tiobe list of languages.}. Due to this imbalance of data, the Code Llama models will be limited in the solutions it could provide for those so-called ``unpopular'' languages, making our results non-representative for this group. Nevertheless, we argue that the popularity is still a good factor for selecting languages as these popular languages are widely used by all kinds of software developers. 

Since Code Llama provides the possibility to change the randomness of the generated solutions by adjusting the temperature of the models, the choice was made to also compare the code generated for three different temperatures: 0.6, 0.8, and 1.0. As described in Section~\ref{c:planning}, we selected these specific temperatures based on initial experiments. Some could consider the difference between these temperatures negligible; however, even these small differences affect the results generated by Code Llama. 

\subsection{Construct Validity}\label{threat:construct}

All three of the benchmark problems used in the experiment depend on the provided input. Both \acrshort{cn} and \acrshort{ts} require an input array, while a file and strings are needed for \acrshort{sr}. The input of these problems is known to affect the performance of the code solutions. Since this study only considers one input per problem, this poses a threat to construct validity. For each of the algorithms, a worst-case scenario input was entered. For \acrshort{sr}, a large text file repeating generic ``lorem ipsum'' text was selected, totalling 268,435,456 characters (the maximum string length in JavaScript). For \acrshort{ts}, a file of 100,000 numbers was generated, ensuring that the answer is represented by the last two values, forcing the algorithm to traverse the entire input. For \acrshort{cn}, 39,000 numbers were inputted, in order to prevent the runtime of the less efficient implementations from ballooning too large.

\subsection{Conclusion Validity}\label{threat:conclusion}
When conducting the experiment, a conscious choice was made to utilise 30 runs to ensure a sufficient sample size. Nevertheless, due to the non-normal distribution of some of the data samples, it was not possible to use the \textit{t-test} for RQ1 and RQ2 nor the \textit{ANOVA} test for RQ3. Therefore, we chose the \textit{Mann-Whitney} for the former and the \textit{Kruskall-Wallis} test for the latter instead.
The results show that the energy efficiency of Code Llama-generated code is heavily influenced by the specific programming languages utilised, as well the specific benchmark problem. While the p-values presented in \autoref{tab:pvalues-1} all indicate rejection of the null hypothesis of RQ1, the results for the statistical tests conducted for RQ2 and RQ3 are conflicting. Therefore, the conclusions with respect to these research questions are more specific to the individual cases and thus pose a threat to conclusion validity. However, these results provide important insights into the influence of the programming language and prompt on the energy efficiency of code generated by Code Llama.

\section{Related Work}\label{c:related}

Several studies have analysed AI-generated code, investigating security \cite{Khoury_2023, Pearce_2021}, correctness \cite{refining-chatgpt-generated-code}, code quality \cite{refining-chatgpt-generated-code}, runtime performance \cite{erhabor2023measuring}, among others. To the best of our knowledge, however, there are no works targeting the energy consumption of AI-generated code. This section presents an overview of the related work in this field.

Khoury~\etal~\cite{Khoury_2023} address the potential for vulnerabilities in the code generated by ChatGPT, through an experiment that systematically evaluates the security of programs generated by the AI chatbot. The experiment involved ChatGPT generating 21 programs in five different programming languages, followed by an evaluation of the resulting code. The findings suggest that ChatGPT is able to identify potential vulnerabilities if prompted, but often provides insecure code. The experiment presented in this study differs from the work by Khoury~\etal in its objective and scope. Instead of considering the security of ChatGPT-generated code, this experiment is focused on the energy efficiency of code produced by the \acrshort{llama} model.

The research conducted by Pearce~\etal~\cite{Pearce_2021} is closely related to \cite{Khoury_2023} as the aim of the paper is to assess the security of code generated by GitHub Copilot. The authors established 89 distinct scenarios by considering the 2021 ``Top 25'' Common Weakness Enumeration (CWE) list established by MITRE, the different context possibilities, and three distinct programming languages: Python, C, and Verilog. Since GitHub Copilot can provide any number of code options, all these options are included, resulting in 1,689 programs. To assess the vulnerability of the generated code, both CodeQL software and manual inspection were utilised. The results indicate that a considerable number of the programs (41\%) completed by Copilot were assessed as vulnerable. Similar to Pearce~\etal, this research considers multiple code options as provided by \acrshort{llama}; however, the code will be analysed on the energy efficiency instead of possible code vulnerabilities.

Liu~\etal~\cite{refining-chatgpt-generated-code} investigate the quality of the code produced by ChatGPT, with regards to both its code generation performance, in terms of functional correctness, and the quality of the generated code. They achieve this by prompting ChatGPT3.5-March23 to generate solutions to 2,033 LeetCode questions, in both Python and Java. The produced outputs are then tested for proper functionality, using unit tests, and for quality, using code linters. The authors found that 66\% of the Python solutions and 69\% of the Java solutions were functionally correct following the first prompt. However, the authors noted that the model performance dropped nearly five-fold on problems released after the dataset was collected, hinting at the fact that some of the prior questions might have been contained within the training set, offering the model an advantage. In terms of quality, 53\% of the Java programs and 35\% of the Python programs contained code style issues, code smells, or similar bad practices. An additional finding is that the model was able to resolve between 20\% and 60\% of issues itself when asked to do so. This work differs from this paper in two main aspects. Firstly, \acrshort{llama} is used in place of ChatGPT for code generation. Secondly, while \cite{refining-chatgpt-generated-code} investigates the correctness of the programs and the overall quality of the code, the present paper investigates the energy efficiency of the resulting algorithms.

Erhabor \etal~\cite{erhabor2023measuring} evaluate Github Copilot in terms of runtime performance. It presents a user study involving 32 participants who were tasked with solving two C++ programming problems, one with the assistance of Copilot and the other without it. The results indicate that using Copilot may lead to significantly slower runtime performance. This related paper also highlights the potential trade-off between using Copilot for code generation, which may improve developers' productivity but has limitations when it comes to runtime performance. While \cite{erhabor2023measuring} primarily focuses on runtime performance, this study examines the impact of \acrshort{llama}'s generated code on energy efficiency.

For the sake of understanding energy efficiency in algorithms, this paper references the article by Albers \cite{albers2010energy} which describes many different facets and methodologies in algorithmic sustainability. In this paper, Albers lays out many techniques for energy-efficient algorithms beginning with power-down state, systems with three to n states, scheduling, network topology, and more. The intention with referencing this paper is to be able to perform a white-box analysis on the \acrshort{llama}-generated code with the goal of spotting and cataloguing the energy-saving techniques that the LLM uses. Should the implementation match one of the more well-defined techniques, this paper will be able to further model the expected energy efficiency in a more formal way.

\section{Conclusions}\label{c:conclusions}   

This report presents the initial results of the energy efficiency of Code Llama-generated code. We analysed the energy consumption of Code Llama code solutions for three specific benchmark problems in three distinct programming languages. The energy usage of the individual solutions were compared first with the corresponding human implementations, then with the Code Llama solutions that were retrieved using a different prompt or a distinct temperature level. We performed statistical analysis on the data obtained through this method to determine the effect that prompt engineering for energy efficiency has on said efficiency. However, several temperature values and prompts that explicitly ask for an energy-efficient version of the implementation do not affect the energy efficiency of the generated code. Software developers and engineers will find these results useful for making decisions about trusting Code Llama-generated code in settings which require efficient algorithms. Moreover, this study hightlights the needs of energy-aware or quality-aware LLMs models. As a result, LLMs models that can optimize and generate high-quality code are required, considering the programming problem and a candidate hardware platform.

Given the scope of the experiments outlined and performed in this paper, our contribution to the field of energy measurement of AI-generated code consists of an initial analysis of Code Llama-generated code. This study could be further improved with a larger set of data; \ie more algorithms ran in additional languages which we were not able to include in this paper. Furthermore, as mentioned before, this paper has been limited by only being able to run the smallest 7B model of Code Llama \cite{codeLLamaModel}. Given more resources, the research would be further improved by generating code with the other available models, such as the larger instruction-based models, the code-completion models, or the python-specific models. One possible limitation of this extension is the amount of time required to do 30 runs of a single algorithm on the Monsoon Power Monitor and the \acrshort{rpi} devices. To overcome this, we suggest 
devices with increased computing power (\eg including multiple GPUs). Future research could include other quality attributes into the analysis, such as performance (\eg memory usage and execution time) to evaluate the capabilities of Code Llama from more perspectives. 
%


\balance

\bibliographystyle{IEEEtran}
\bibliography{references}

\begin{thebibliography}{10}
\providecommand{\url}[1]{#1}
\csname url@samestyle\endcsname
\providecommand{\newblock}{\relax}
\providecommand{\bibinfo}[2]{#2}
\providecommand{\BIBentrySTDinterwordspacing}{\spaceskip=0pt\relax}
\providecommand{\BIBentryALTinterwordstretchfactor}{4}
\providecommand{\BIBentryALTinterwordspacing}{\spaceskip=\fontdimen2\font plus
\BIBentryALTinterwordstretchfactor\fontdimen3\font minus \fontdimen4\font\relax}
\providecommand{\BIBforeignlanguage}[2]{{%
\expandafter\ifx\csname l@#1\endcsname\relax
\typeout{** WARNING: IEEEtran.bst: No hyphenation pattern has been}%
\typeout{** loaded for the language `#1'. Using the pattern for}%
\typeout{** the default language instead.}%
\else
\language=\csname l@#1\endcsname
\fi
#2}}
\providecommand{\BIBdecl}{\relax}
\BIBdecl

\bibitem{Yepis_2023}
\BIBentryALTinterwordspacing
E.~Yepis, ``Hype or not? ai’s benefits for developers explored in the 2023 developer survey,'' Jun 2023. [Online]. Available: \url{https://stackoverflow.blog/2023/06/14/hype-or-not-developers-have-something-to-say-about-ai/}
\BIBentrySTDinterwordspacing

\bibitem{Huang_2023}
K.~Huang, Z.~Xu, S.~Yang, H.~Sun, X.~Li, Z.~Yan, and Y.~Zhang, ``A survey on automated program repair techniques,'' 2023.

\bibitem{Touvron_2023}
H.~Touvron, T.~Lavril, G.~Izacard, X.~Martinet, M.-A. Lachaux, T.~Lacroix, B.~Rozière, N.~Goyal, E.~Hambro, F.~Azhar, A.~Rodriguez, A.~Joulin, E.~Grave, and G.~Lample, ``Llama: Open and efficient foundation language models,'' 2023.

\bibitem{Roziere_2023}
B.~Rozière, J.~Gehring, F.~Gloeckle, S.~Sootla, I.~Gat, X.~E. Tan, Y.~Adi, J.~Liu, T.~Remez, J.~Rapin, A.~Kozhevnikov, I.~Evtimov, J.~Bitton, M.~Bhatt, C.~C. Ferrer, A.~Grattafiori, W.~Xiong, A.~Défossez, J.~Copet, F.~Azhar, H.~Touvron, L.~Martin, N.~Usunier, T.~Scialom, and G.~Synnaeve, ``Code llama: Open foundation models for code,'' 2023.

\bibitem{perry2022users}
N.~Perry, M.~Srivastava, D.~Kumar, and D.~Boneh, ``Do users write more insecure code with ai assistants?'' 2022.

\bibitem{Khoury_2023}
R.~Khoury, A.~R. Avila, J.~Brunelle, and B.~M. Camara, ``How secure is code generated by chatgpt?'' 2023.

\bibitem{Pearce_2021}
H.~Pearce, B.~Ahmad, B.~Tan, B.~Dolan-Gavitt, and R.~Karri, ``Asleep at the keyboard? assessing the security of github copilot's code contributions,'' 2021.

\bibitem{Wang_2022}
\BIBentryALTinterwordspacing
P.~Wang, P.~Zhong, M.~Yu, Y.~Pu, S.~Zhang, and P.~Yu, ``Trends in energy consumption under the multi-stage development of ict: Evidence in china from 2001 to 2030,'' \emph{Energy Reports}, vol.~8, pp. 8981--8995, 2022. [Online]. Available: \url{https://www.sciencedirect.com/science/article/pii/S2352484722012677}
\BIBentrySTDinterwordspacing

\bibitem{Singh_2015}
\BIBentryALTinterwordspacing
J.~Singh, K.~Naik, and V.~Mahinthan, ``Impact of developer choices on energy consumption of software on servers,'' \emph{Procedia Computer Science}, vol.~62, pp. 385--394, 2015, proceedings of the 2015 International Conference on Soft Computing and Software Engineering (SCSE'15). [Online]. Available: \url{https://www.sciencedirect.com/science/article/pii/S1877050915025582}
\BIBentrySTDinterwordspacing

\bibitem{replication_package1}
``Replication package of this study: Prompts, experiment scripts, and implementations,'' \url{https://anonymous.4open.science/r/quatic-repl-package-code-llama/README.md}, online.

\bibitem{Basili94}
V.~R. Basili, G.~Caldiera, and D.~H. Rombach, \emph{{T}he {G}oal {Q}uestion {M}etric {A}pproach}.\hskip 1em plus 0.5em minus 0.4em\relax John Wiley \& Sons, 1994, vol.~I.

\bibitem{refining-chatgpt-generated-code}
\BIBentryALTinterwordspacing
Y.~Liu, T.~Le{-}Cong, R.~Widyasari, C.~Tantithamthavorn, L.~Li, X.~D. Le, and D.~Lo, ``Refining chatgpt-generated code: Characterizing and mitigating code quality issues,'' \emph{CoRR}, vol. abs/2307.12596, 2023. [Online]. Available: \url{https://doi.org/10.48550/arXiv.2307.12596}
\BIBentrySTDinterwordspacing

\bibitem{Amazon_CodeLLama_2023}
\BIBentryALTinterwordspacing
Amazon, ``Code llama code generation models from meta are now available via amazon sagemaker jumpstart,'' October 2023. [Online]. Available: \url{https://aws.amazon.com/blogs/machine-learning/code-llama-code-generation-models-from-meta-are-now-available-via-amazon-sagemaker-jumpstart/}
\BIBentrySTDinterwordspacing

\bibitem{etinski2012understanding}
M.~Etinski, J.~Corbal{\'a}n, J.~Labarta, and M.~Valero, ``Understanding the future of energy-performance trade-off via dvfs in hpc environments,'' \emph{Journal of Parallel and Distributed Computing}, vol.~72, no.~4, pp. 579--590, 2012.

\bibitem{Pereira2017}
\BIBentryALTinterwordspacing
R.~Pereira, M.~Couto, F.~Ribeiro, R.~Rua, J.~Cunha, J.~a.~P. Fernandes, and J.~a. Saraiva, ``Energy efficiency across programming languages: How do energy, time, and memory relate?'' in \emph{Proceedings of the 10th ACM SIGPLAN International Conference on Software Language Engineering}, ser. SLE 2017.\hskip 1em plus 0.5em minus 0.4em\relax New York, NY, USA: Association for Computing Machinery, 2017, p. 256–267. [Online]. Available: \url{https://doi.org/10.1145/3136014.3136031}
\BIBentrySTDinterwordspacing

\bibitem{monsoon_2022}
{Monsoon Solutions}, ``Monsoon power monitor,'' \url{https://www.msoon.com/}, accessed: 2021-09-26.

\bibitem{codeLLamaModel}
Software and V.~U.~A. Sustainability~Group, ``{Code Llama: Open Foundation Models for Code},'' \url{https://github.com/S2-group/experiment-runner}, 2023.

\bibitem{zhang2022opt}
S.~Zhang, S.~Roller, N.~Goyal, M.~Artetxe, M.~Chen, S.~Chen, C.~Dewan, M.~Diab, X.~Li, X.~V. Lin \emph{et~al.}, ``Opt: Open pre-trained transformer language models,'' \emph{arXiv preprint arXiv:2205.01068}, 2022.

\bibitem{experimentRunner}
Software and V.~U.~A. Sustainability~Group, ``{Experiment Runner: A Framework to automatically execute measurement-based experiments on any platform},'' \url{https://github.com/S2-group/experiment-runner}, 2023.

\bibitem{macbeth2011cliff}
G.~Macbeth, E.~Razumiejczyk, and R.~D. Ledesma, ``Cliff's delta calculator: A non-parametric effect size program for two groups of observations,'' \emph{Universitas Psychologica}, vol.~10, no.~2, pp. 545--555, 2011.

\bibitem{verdecchia2021green}
R.~Verdecchia, P.~Lago, C.~Ebert, and C.~De~Vries, ``Green it and green software,'' \emph{IEEE Software}, vol.~38, no.~6, pp. 7--15, 2021.

\bibitem{bilsen2022study}
V.~Bilsen, J.~Gr{\"o}ger, W.~Devriendt, R.~Liu, S.~Gau{\v{s}}as, F.~Behrens, F.~Bley, M.~Carpentier, V.~Duch{\^e}ne, A.~R. K{\"o}hler \emph{et~al.}, ``Study on greening cloud computing and electronic communications services and networks,'' 2022.

\bibitem{Cook:1979}
T.~Cook and D.~Campbell, \emph{\BIBforeignlanguage{English}{Quasi-Experimentation: Design and Analysis Issues for Field Settings}}.\hskip 1em plus 0.5em minus 0.4em\relax Houghton Mifflin, 1979.

\bibitem{erhabor2023measuring}
D.~Erhabor, S.~Udayashankar, M.~Nagappan, and S.~Al-Kiswany, ``Measuring the runtime performance of code produced with github copilot,'' \emph{arXiv preprint arXiv:2305.06439}, 2023.

\bibitem{albers2010energy}
S.~Albers, ``Energy-efficient algorithms,'' \emph{Communications of the ACM}, vol.~53, no.~5, pp. 86--96, 2010.

\end{thebibliography}

\end{document}